\documentclass{article}
\usepackage{amssymb}
\usepackage{graphicx}
\begin{document}
\def\boxit#1{\vcenter{\hrule\hbox{\vrule\kern8pt
      \vbox{\kern8pt#1\kern8pt}\kern8pt\vrule}\hrule}}
\def\Boxed#1{\boxit{\hbox{$\displaystyle{#1}$}}} 
\def\sqr#1#2{{\vcenter{\vbox{\hrule height.#2pt
        \hbox{\vrule width.#2pt height#1pt \kern#1pt
          \vrule width.#2pt}
        \hrule height.#2pt}}}}
\def\square{\mathchoice\sqr34\sqr34\sqr{2.1}3\sqr{1.5}3}
\def\Square{\mathchoice\sqr67\sqr67\sqr{5.1}3\sqr{1.5}3}
\def\lambdabar{{\mathchar'26\mkern-9mu\lambda}}
\def\thrdotovervx{\buildrel\textstyle...\over v_x}
\def\thrdotovervy{\buildrel\textstyle...\over v_y}
\title{\bf Zitterbewegung in Noncommutative Geometry }
\author{{\small \ Mehran Zahiri Abyaneh}\footnote{me\_zahiri@sbu.ac.ir} \ {\small and\
         Mehrdad Farhoudi}\footnote{m-farhoudi@sbu.ac.ir}\\
        {\small Department of Physics, Shahid Beheshti University, G.C.,
        Evin, Tehran 19839, Iran}}
\date{\small March 31, 2019}
\maketitle
\begin{abstract}
\noindent
 We have considered the effects of space and momentum
noncommutativity separately on the \emph{zitterbewegung}\ ({\bf
ZBW}) phenomenon. In the space noncommutativity scenario, it has
been expressed that, due to the conservation of momentum, the
Fourier decomposition of the expectation value of position
does~not change. However, the noncommutative ({\bf NC}) space
corrections to the magnetic dipole moment of electron, that was
traditionally perceived to come into play only in the first-order
of perturbation theory, appear in the leading-order calculations
with the similar structure and numerically the same order, but
with an opposite sign. This result may explain why for large lumps
of masses, the Zeeman-effect due to the noncommutativity remains
undetectable. Moreover, we have shown that the $x$- and
$y$-components of the electron magnetic dipole moment, contrary to
the commutative (usual) version, are non-zero and with the same
structure as the $z$-component. In the momentum noncommutativity
case, we have indicated that, due to the relevant external uniform
magnetic field, the energy-spectrum and also the solutions of the
Dirac equation are changed in $3+1$ dimensions. In addition, our
analysis shows that in $2+1$ dimensions, the resulted NC field
makes electrons in the zero Landau-level rotate not only via a
cyclotron motion, but also through the ZBW motion with a frequency
proportional to the field, which doubles the amplitude of the
rotation. In fact, this is a hallmark of the ZBW in graphene that
provides a promising way to be tested experimentally.
\end{abstract}
\bigskip
\section{Introduction}
\indent

It is well-known that for a free Dirac
particle~\cite{dirac2-1928}\ the velocity and momentum do~not
coincide, that is, a free particle oscillates rapidly with the
speed $c$ around the center of mass while moving like a
relativistic particle with velocity ${\bf
p}/m$~\cite{Dirac58}--\cite{Merzbacher98}. This rapid oscillatory
motion was called \emph{zitterbewegung} -- a trembling/quivering
motion -- by Schr\"odinger~\cite{schrodinger30-31,barut-etal81a}.
The amplitude of this motion is predicted to be of the order of
the Compton wavelength for electrons, i.e.
$\lambdabar_c=\hbar/(m_ec)\simeq 3.86\times {10}^{-13}\, {\rm m}$.
Using this picture, it has been suggested in the literature that
the spin and magnetic moment of an electron, as a point charge, is
generated by an intrinsic local
motion~\cite{huang1952}--\cite{hestenes2009}. In other words, the
magnetic moment and the spin of electron are consequences of a
local circular motion of mass and charge of electron, and may be
considered as an ``orbital angular momentum" due to the ZBW. Other
attempts at explaining the spin of
electron~\cite{Belinfante,Ohanian} have also established that it
can be regarded as due to a circulating flow of energy in the same
footnote as orbital angular momentum. Similarly, in
Ref.~\cite{Sasabe2008}, it has been stated that the origin of the
spin-magnetic moment is caused by a quantum transition current
between positive and negative energy states of the solutions of
the Dirac equation, and hence, it relates closely to the ZBW
phenomenon. Indeed, not only the Dirac strong
support~\cite{Dirac58} of Schr\"odinger's ZBW as a fundamental
property of electron has been unchallenged up today, but there is
increasing evidence that ZBW is a real effect, in principle,
observable, e.g., in a Bose-Einstein condensate~\cite{heszitter2}
and in semiconductors~\cite{Zawadzki2011}. Nevertheless, by a
unitary Foldy-Wouthuysen transformation~\cite{Foldy}, the ZBW can
be avoided because this transformation actually eliminates
negative energy components in electron wave functions. However,
the ZBW goes hand in hand with the existence of negative energy
solutions, and is only important for wave packets with significant
interference between positive and negative frequencies. Even in
this regard, taking the neutrino as a localized wave packet
consisting of positive and negative energies, and studying its
chiral ZBW, it has been claimed~\cite{Stefano} that it may explain
the ``missing'' solar neutrino experiments, and also
interpreting~\cite{Bernardini} chiral oscillations in terms of the
ZBW effect. Moreover, the results of such that transformation are
in contrast with the Darwin term~\cite{Darwin} in atomic physics.

Furthermore, in the issue of ZBW, in the approach of
Ref.~\cite{huang1952}, it has been demonstrated that for a wave
packet consisting of both positive and negative energy components,
with the specified initial condition, the contribution of each
plane wave to the motion of the wave packet is an orbit whose
projection on the plane perpendicular to the direction of the spin
is a circle of radius $\lambdabar_c/2$ with frequency $\omega^{\rm
[zbw]}(\simeq 1.6\times {10}^{21}\, {\rm s}^{-1}$ for electrons).
This result leads to the intrinsic magnetic moment of particle
with the correct gyromagnetic $g$ factor. We also
studied~\cite{zahirifarhoudi} the same issue, but in the presence
of an external magnetic field, and showed that the previous
results are stable and the only difference is that in this case
the ZBW frequency and the amplitude are shifted.

On the other hand, some scientists believe that the NC geometry
can give new vision for the spacetime structure especially nearly
at the Planck scale, see, e.g.,
Refs.~\cite{MF16}--\cite{SabaFarhoudib}. In this respect, there is
a vast literature discussion on the theoretical and
phenomenological consequences of these effects, see, e.g.,
Refs.~\cite{Douglas}--\cite{Hinchliffe} and references therein.
The noncommutativity concept between spacetime coordinates was
first introduced in 1947~\cite{snyder}, and thereafter, the NC
geometry has taken shape since 1980, see, e.g.,
Refs.~\cite{Connes}--\cite{MF12} and references therein. This
picture has also strong motivation in the framework of the string
and M-theories, see, e.g.,
Refs.~\cite{SeibergWitten}--\cite{ADNS}. Indeed, it has been
claimed~\cite{Singh2005} that as standard linear quantum mechanics
should be a limiting case of an underlying new non-linear quantum
theory, a possible approach for such a new formulation can be
sought through the use of NC geometry. In this regard, due to
changes raised in this picture, quantum mechanics and quantum
field theory ({\bf QFT}) phenomena have been affected and some
research have been devoted to this
sector~\cite{Gamboa}--\cite{Cai}. Actually, while considering the
NC geometry, the QFT modifies the relativistic wave equation, i.e.
the Dirac equation, by which, one can calculate the consequences
of NC parameters on different physical effects. However, there
have been two approaches in including the NC effects on the Dirac
equation for an electron in an external electromagnetic field,
namely just by the simple Moyal modification, and another one,
this modification plus taking the so called Seiberg-Witten ({\bf
SW}) map also into account~\cite{SeibergWitten}, wherein, it has
been shown that only the latter one keeps the Dirac equation being
gauge invariant~\cite{Adorno,bertolami}.

Now, regarding the ZBW and its relation with the electron magnetic
dipole moment, it is intriguing to figure out effect(s) of the NC
geometry on the ZBW phenomenon, and indeed, on the internal
structure of electron and its spin as well. However, as the scale
of the ZBW is lower than practical experimental resolution and
since the NC effects are also small, there may be some claims that
such effects are hardly being detected in the near future.
Nevertheless, and interestingly enough, the phenomenon of ZBW for
electrons has been shown (e.g.,
Refs.~\cite{Zawadzki2011,Ferrari,Castro} and other references
mentioned in Refs.~\cite{zahirifarhoudi,Sasabe2014}) to occur in
non-relativistic cases and in two-dimensional Dirac materials
(like graphene), wherein the ZBW has much lower angular frequency
and much larger amplitude and thus, it may lend itself to
experimental detection easier. On the other hand, the effect of
the NC geometry has been studied, e.g.
Refs.~\cite{Dayi}--\cite{Santos1}, in two-dimensional Dirac
materials and on the quantum Hall effect as well.

Having all these motivations into account, the purpose of this
work is to investigate effect(s) of the NC geometry on the ZBW
phenomenon; and to perform this task, at the beginning we
concisely review the ZBW in the usual commutative geometry in
$3+1$ dimensions. Afterward, we continue with a brief introduction
on the noncommutativity and how it appears in the Dirac equation
while giving an introduction to the basic premises of the NC
geometry. Then, we first consider the effect of space
noncommutativity on the ZBW phenomenon and on the magnetic dipole
moment of an electron in the leading-order calculations in $3+1$
dimensions. In Sec.~$4$, we first continue to investigate the
effect of momentum noncommutativity on the Dirac Hamiltonian and
its solutions in $3+1$ dimensions, and thereupon, consider the
issue in the $2+1$ dimensional case for graphene. Finally, we
conclude the summary of the results in the last section.

\section{ ZBW Phenomenon in Commutative Geometry}
\indent

In this section, we consider the ZBW for a localized wave packet
consisting of positive and negative energies, as in
Ref.~\cite{huang1952}, with some necessary explanations as a
calculation in the commutative (usual) version. Thereafter, in the
next section, we carry out the same approach for the NC situation.
Thus, we will be able to figure out the effect(s) of the NC
geometry when it is compared with the results of the commutative
one.

In this regard, the Dirac equation for an electron in a field-free
region is
\begin{equation}\label{EE}
i\hbar\partial_{t}\Psi=\left(c\,\mbox{\boldmath$\alpha$}\cdot{\bf
p} +\gamma_0 m_ec^2\right)\Psi ,
\end{equation}
wherein the matrices $\mbox{\boldmath$\alpha$}$ and $\gamma_0$ are
defined as\footnote{In terms of the Dirac (gamma) matrices,
$\alpha_i\equiv\gamma^0\gamma_i$ for $i=1, 2, 3$.}
\begin{eqnarray}\label{DD}
\alpha_i=\left(
                 \begin{array}{cc}
                   0 & \sigma_i \\
                  \sigma_i & 0 \\
                 \end{array}
               \right)
               \qquad\quad\ \textrm{and}\qquad\quad\ \gamma_0=\left(
                 \begin{array}{cc}
                   I & 0 \\
                   0 & -I \\
                 \end{array}
               \right),
\end{eqnarray}
where the Latin lowercase letters run from one to three, and
$\sigma_i$ and $I$ are the $2\times 2$ Pauli matrices and the unit
matrix, respectively. The solution of Eq. (\ref{EE}) can be
written as a general wave packet expanded in terms of momentum
eigenfunctions as~\cite{huang1952}
\begin{equation}\label{FF}
\!\!\!\!\!\Psi({\bf
r},t)=h^{-3/2}\int{\Bigl[C_+({\mathbf{p}})\exp(-i\omega t)
+C_{-}({\mathbf{p}})\exp(i\omega
t)\Bigr]\exp\Bigl(i\frac{{\mathbf{p}}\cdot{\bf r}}{\hbar}\Bigr)
d^{3}p},
\end{equation}
where $\omega =\gamma m_e c^2/\hbar$ and $C_{+}({\mathbf{p}})$\
$\left(C_{-}({\mathbf{p}})\right)$ is a linear combination of the
spin-up and spin-down amplitudes of the free particle Dirac waves
with momentum $\textbf{p}$ and positive (negative) energy. Such a
wave packet includes both negative and positive energies. This
wave packet can be used with an initial condition which being a
localized spin-up electron in the $z$-direction while its center
is at rest in the origin, namely
\begin{equation}\label{GG}
\Psi({\bf r},0)=\left(
            \begin{array}{c}
            1 \\
            0 \\
            0 \\
            0\\
            \end{array}
          \right)f\left(\frac{r}{r_{_o}}\right).
\end{equation}
Here, $f(r/r_{_o})=\left[2/(\pi
r_{_o}^2)\right]^{3/4}\exp\left[-(r/r_{_o})^2\right]$ is a
normalized Gaussian function with $r\equiv|{\bf r}|$ and
${r_{_o}}$ as a constant that indicates approximate spatial
extension of the wave packet. Also, the Fourier transformation of
$f(r/r_{_o})$ is
\begin{equation}\label{HH}
f\left(\frac{p}{p_{_{o}}}\right)=\left(\frac{2}{\pi
p_{_o}^2}\right)^{3/4}\exp\left[-\left(\frac{p}{p_{_o}}\right)^2\right],
\end{equation}
that satisfies the normalization condition $\int
f^2(p/p_{_{o}})\,d^3 p=1$, and where the constant
$p_{_{o}}=2\hbar/r_{_{o}}$ gives the width of wave packet in the
momentum space~\cite{huang1952}. Now, to simplify the model
further, if one applies the non-relativistic approximation and
hence drops the terms of the order $(p/2m_ec)^2$ or higher, then
in this case, the wave packet~(\ref{FF}) will read
\begin{equation}\label{II}
\Psi({\bf r},t)\simeq h^{-3/2}\int\Biggl\{\left(
                 \begin{array}{c}
                   1 \\
                 0 \\
                 K p_3 \\
                 K p_+ \\
                 \end{array}
               \right)\exp(-i\omega t)+\left(
               \begin{array}{c}
                 0 \\
               0 \\
               -K p_3 \\
               -K p_+\\
               \end{array}
             \right)\exp(i\omega t)
             \Biggr\}f(p/p_{_{o}})
            \exp\Bigl(i\frac{{\mathbf{p}}\cdot{\bf r}}{\hbar}\Bigr) \,d^3 p ,
\end{equation}
where $\omega \simeq m_e c^2/\hbar$ in the non-relativistic
approximation, $p_+=p_1+ip_2$ and $K \simeq\frac{1}{2m_ec}$. The
above rough solution satisfies the Dirac equation (\ref{EE}) when
the aforementioned approximation is taken into account.

Using this wave packet and the expectation value of velocity
vector, i.e. $ <\dot{\bf r}>=\int\Psi^{\ast}({\bf
r},t)(c\mbox{\boldmath$\alpha$}) \Psi({\bf r},t)d^{3}r$ in the
spherical coordinates, in order to specify the Fourier
decompositions, one gets~\cite{huang1952,zahirifarhoudi}
\begin{eqnarray}\label{JJ}
&<x>\simeq
I\frac{\lambdabar_c}{2}\int_{0}^{2\pi}\sin\left(\omega^{\rm
[zbw]}t+\varphi\right)\, d\varphi ,\qquad\quad
 <y>\simeq
I\frac{\lambdabar_c}{2}\int_{0}^{2\pi}\cos\left(\omega^{\rm
[zbw]}t+\varphi\right)\, d\varphi\cr {}\cr
 &
{\rm and}\qquad\quad <z>\simeq J\lambdabar_c\sin\left(\omega^{\rm
[zbw]}t\right),
\end{eqnarray}
where $\omega^{\rm [zbw]}\equiv 2\,\omega$ in the non-relativistic
approximation, $\varphi $ is the azimuthal angle in the spherical
momentum space, and
\begin{equation}\label{KK}
I\equiv-2\!\!\int_{0}^{\infty}\!\int_{0}^{\pi}\frac{f^2(p/p_{_{o}})}{2m_ec}
p^3\sin^2\theta\,d\theta
\,dp=-(8\pi)^{-\frac{1}{2}}\frac{\lambdabar_c}{r_{_o}}
\end{equation}
and
\begin{eqnarray}\label{LL}
J\equiv-\pi\int_{0}^{\infty}\int_{0}^{\pi}\frac{f^2(p/p_{_{o}})}{2m_ec}
p^3 \sin2\theta\,d\theta\,dp=0.
\end{eqnarray}
Of course, all components of $<\!{\bf r}\!>$ vanish upon
integration over the full domain, and thus all oscillatory motions
disappear while the center of the wave packet as a whole remains
at rest. However, to realize the role of a specific Fourier
decomposition in the ZBW circular motion, one can confine oneself
to a fixed value of $\varphi $, say $\varphi_{_{o}}$. Thus in this
case, the Fourier components in the $xy$-plane
become~\cite{huang1952,zahirifarhoudi}
\begin{equation}\label{Spc11}
<x>_{\varphi_{_{o}}}\simeq
\frac{\lambdabar_c}{2}\sin\left(\omega^{\rm
[zbw]}t+\varphi_{_{o}}\right)\qquad\quad {\rm and}\qquad\quad
 <y>_{\varphi_{_{o}}}\simeq
\frac{\lambdabar_c}{2}\cos\left(\omega^{\rm
[zbw]}t+\varphi_{_{o}}\right),
\end{equation}
i.e., a circle of radius $\lambdabar_c/2$ with the frequency
$\omega^{\rm [zbw]}$. However, the consequence of the ZBW is that
it is impossible to localize electron better than to a certain
finite volume, as its weight relative to the degree of
localization of electron in space, given by $I$ in
definition~(\ref{KK}), is proportional to $\lambdabar_c/r_{_o}$.

As the expectation value of the magnetic moment of a spin-up/down
electron with charge $e<0$, in the momentum representation
operator form in the Gaussian unit, is
\begin{equation}\label{MM}
<\!\mbox{\boldmath$\mu$}\!>=\frac{e}{2c}<\!{\bf r}\times\dot{\bf
r}\!> =\frac{e}{2}<\!{\bf
r}\times\mbox{\boldmath$\alpha$}\!>\rightarrow
\frac{ie\hbar}{2}<\!{\mbox{\boldmath$\nabla$}}_{\bf
p}\times\mbox{\boldmath$\alpha$}\!>,
\end{equation}
straightforward calculations, using the wave packet (\ref{II}) and
the employed approximations,
reveal~\cite{huang1952,zahirifarhoudi}
\begin{equation}\label{NN}
<\!\mu^\uparrow_1\!>=0,\qquad\quad
<\!\mu^\uparrow_2\!>=0\qquad\quad {\rm and}\qquad\quad
<\!\mu^\uparrow_3\!>=\frac{ e\lambdabar_c}{2
}\left[1-\cos(\omega^{\rm [zbw]}t)\right]
\end{equation}
for the spin-up states. Also in the same manner, while employing
the relevant wave packet, one correspondingly achieves
\begin{equation}\label{OO}
<\!\mu^\downarrow_1\!>=0,\qquad\quad
<\!\mu^\downarrow_2\!>=0\qquad\quad {\rm and}\qquad\quad
<\!\mu^\downarrow_3\!>=-\frac{e\lambdabar_c}{2
}\left[1-\cos(\omega^{\rm [zbw]}t)\right]
\end{equation}
for the spin-down states, wherein the same initial condition as
(\ref{GG}) but with a localized spin-down electron, namely
\begin{equation}\label{bbb}
\Psi({\mathbf{r}},0)=\left(
            \begin{array}{c}
            0 \\
            1 \\
            0 \\
            0\\
            \end{array}
          \right)f\left(\frac{r}{r_{_o}}\right),
\end{equation}
has been chosen to perform the rest of calculations. Note that,
the above $x$- and $y$-components not only vanish on the average,
but also their Fourier components vanish individually. Thus, it
supports that each Fourier wave contributes a circular motion
about the direction of the spin, wherein the $z$-component of
(\ref{MM}) is the only non-vanishing component of the magnetic
moment. Hence, one may interpret that the intrinsic magnetic
moment is a result of the ZBW.

We should mention that the electron magnetic moment has also been
derived in Ref.~\cite{Sasabe2014} via the role of ZBW for a
spin-up electron, while using the Heisenberg picture, and has been
arrived with the same result for the $z$-component as
relation~(\ref{NN}), but without the time-dependent part. However,
by performing the calculations for the $x$- and $y$-components of
the magnetic moment through the approach mentioned in
Ref.~\cite{Sasabe2014}, the results show that these components
do~not vanish. Incidentally, the time-dependant part of the
magnetic moment, in relations~(\ref{NN}) and~(\ref{OO}), is a
consequence of the fact that one has assumed that the spin of
electron had initially been observed. This requirement leads to
bring in the negative energy-part which does~not vanish by letting
the wave packet spread in space~\cite{huang1952,zahirifarhoudi}.

\section{Effect of Space Noncommutativity on ZBW}
\indent

The noncommutativity of physical quantities is one of the most
important peculiarities of quantum mechanics, wherein the usual
fundamental algebra among the momentum and position operators,
namely the commutators $[x_i, x_j]=0$, $[p_i, p_j]=0$ and $[x_i,
p_j]=i\hbar\delta_{ij}$, are generalized in the case of NC
phase-space as
\begin{equation}\label{AA}
[x_{i}, x_{j}] = i \theta_{ij},\qquad\quad [p_{i}, p_{j}] = i
\eta_{ij}\qquad\quad {\rm and}\qquad\quad
[x_i,p_j]=i\hbar\left(\delta_{ij}+{\theta_{ik}\eta_{jk}\over
4\hbar^2}\right).
\end{equation}
The real constant antisymmetric components $\theta_{ij}$ and
$\eta_{ij}$ are the NC parameters of the space and momentum
sectors, with dimensions of length squared and momentum squared,
respectively. In terms of the Levi-Civita antisymmetric tensor,
these parameters can be written as
\begin{equation}\label{BB}
{\theta}_k\equiv {1\over
2}\,\varepsilon_{ijk}\,{\theta}_{ij}\qquad\quad {\rm and}
\qquad\quad {\eta}_k\equiv {1\over
2}\,\varepsilon_{ijk}\,{\eta}_{ij}.
\end{equation}

Actually, in classical physics, the product of arbitrary functions
of noncommuting variables, through the Moyal star-product,
reads~\cite{SeibergWitten}
\begin{equation}\label{B1}
(f*g)(\zeta) =\exp
\Bigl[\frac{i}{2}\alpha^{ab}\partial_a^{(1)}\partial_b^{(2)}
\Bigr]f(\zeta^{(1)})g(\zeta^{(2)}){\Bigr
|}_{\zeta^{(1)}=\zeta=\zeta^{(2)}},
\end{equation}
where $\zeta^a=(x^i,p^j)$, $a, b=1, 2, \cdots, 2n$ and $2n$ is the
dimension of phase-space. The real matrix $(\alpha_{ab})$ is a
generalized symplectic structure and can be written as
\begin{equation}\label{B2}
(\alpha_{ab})=\left(%
\begin{array}{cc}
\theta_{ij} & \delta_{ij}-\theta_{k(i}\eta_{j)l}\frac{\delta^{kl}}{4} \\
-\delta_{ij}+\theta_{k(i}\eta_{j)l}\frac{\delta^{kl}}{4}  &  \eta_{ij}  \\
\end{array}%
\right).
\end{equation}
Using the non-canonical linear transformation
$(x,p)\mapsto(x',p')$, the NC algebra can be mapped into the
commutative form where $[x'_i,x'_j]=0$, $[p'_i,p'_j]=0$ and
$[x'_i,p'_j]=i\hbar\delta_{ij}$. One example of such a map,
corresponded to (\ref{B2}), is
\begin{equation}\label{RR}
x_i=\left(x'_i-{\theta_{ij}\over 2\hbar}p'_j\right)\qquad\quad
{\rm and}\qquad\quad  p_i=\left(p'_i+{\eta_{ij}\over
2\hbar}x'_j\right),
\end{equation}
which is sometimes called Bopp-shift method~\cite{CuFaZa,MaWaYa}.
Furthermore, when one shifts the canonical variables through
(\ref{RR}), the Hamiltonian of a system including the NC variables
are usually assumed to have the same functional form as in the
commutative one, i.e.
\begin{equation}\label{NCHamil}
H^{\rm [NC]}\equiv H(x_i, p_j)= H\left(x'_i-{\theta_{il}\over
2\hbar}p'_l,\, p'_j+{\eta_{jk}\over 2\hbar}x'_k\right).
\end{equation}
However this function is defined on the commutative space, but
obviously, the effects of NC parameters now arise through their
equations of motion. Incidentally, the real parameters $\theta_k$
and ${\eta}_k$ are usually assumed to be very small, and are
considered up to the first-order~\cite{Santos2,farhoudi}.

Hence, in this section, and afterward in the following section, we
just consider the limit wherein only first $\theta_k$, and then
${\eta}_k$, is taken to be non-zero, respectively. We also stick
with the $\theta_{0i}=0$ case, for $\theta_{0i}\neq 0$ renders the
theory to be non-unitary~\cite{Gomis}. However, in the case of
${\eta}_k=0$ and in the field-free limit, but with
\mbox{\boldmath$\theta $}$\neq 0$ (that we are going to consider
in this section), its corresponding Dirac equation will be the
same as Eq.~(\ref{EE}). Thus, one may naively expect that the ZBW
phenomenon is~not affected by the noncommutativity. In fact, this
is the case for results~(\ref{JJ}) that do~not change in this
scenario. Although, in the NC quantum electrodynamics ({\bf QED}),
a phase factor still appears while calculating the Feynman
amplitudes~\cite{RiadS2000}--\cite{Eom2002}, however, in our case,
even this phase vanishes identically due to the conservation of
momentum. Now, let us move forward and perform the same
calculations for the electron magnetic moment as in the previous
section, but in the realm of the NC geometry for the case of this
section, while considering the ZBW phenomenon.

However before we proceed, let us meanwhile remind that the effect
of NC geometry on the magnetic moment of electron, without
considering the ZBW and in the framework of the NC QED, has been
found~\cite{RiadS2000}--\cite{Eom2002} to be null in the ordinary
calculations\rlap,\footnote{Nevertheless, due to the structure of
proton as a non-elementary particle, it has been
shown~\cite{CSjT04} that the hydrogen atom spectrum receives
tree-level correction due to noncommutativity.}\
 while its (total) correction in the
one-loop level, due to the vertex correction diagram, appears (in
the Gaussian unit) as
\begin{equation}\label{UU}
<\!\mbox{\boldmath$\mu$}^{\uparrow\downarrow}\!>_{\rm
tot}^{\rm{[NC]}_{\eta =0}}=\frac{ e\lambdabar_c}{2
}\left[\left(1+\frac{\alpha_{\rm
fsc}}{2\pi}\right)\mbox{\boldmath$\sigma$}_{\uparrow\downarrow}
+\frac{m_e c\alpha_{\rm fsc} \gamma_{_E} }{3\pi
\lambdabar_c^2}\mbox{\boldmath$\theta$}\right],
\end{equation}
where $(1+\alpha_{\rm fsc}/2\pi)$ is the gyration factor,
$\alpha_{\rm fsc}=e^2/(\hbar c) $ is the dimensionless
fine-structure constant and $\gamma_{_E} $ is the Euler constant.
As it is obvious, the extra \mbox{\boldmath$\theta $}-dependant
term is a constant independent of electron kinematical states,
i.e. the spin and momentum. Hence, as the energy difference of the
two states is independent of \mbox{\boldmath$\theta $}, it has
been suggested~\cite{Arfaei} that this independent magnetic moment
can be observed via a Stern-Gerlach apparatus, and not by a spin
resonance experiment. Nevertheless, as this correction is
spin-independent, it has been claimed~\cite{Hinchliffe} that its
effect is~not easy to be observed. On the other hand, the NC
contribution to the Zeeman-effect of the hydrogen atom in the
first-order of perturbation theory has also been
derived~\cite{CSjT01} to be the energy-shift
\begin{equation}
\Delta E_{\rm Zeeman}^{\rm{[NC]}_{\eta =0}}=\frac{e\alpha_{\rm
fsc}\gamma_{_E}}{6\pi \lambdabar_c}\left(1-f\frac{m_p}{m_e}\right)
\mbox{\boldmath$\theta$}\!\cdot\!{\bf B},
\end{equation}
where, as proton is~not point-like, a form factor $f$ has been
used which is of the order of unity. Now, as the NC contribution
does~not depend on the direction of the spin, this energy-shift
leads to a cumulative contribution from each atom. Hence, as the
ratio $ f m_p/m_e\gg 1 $, its value has to be an enormous amount
for large lumps of masses (like planets and stars), unless the
value of $\mbox{\boldmath$\theta$}$ is infinitesimally
small\footnote{Indeed, the Lamb-shift data, the $e^+e^-$
scattering data and some other experiments have been used to
impose some bounds on the value of the NC parameter
$\mbox{\boldmath$\theta$}$, see, e.g., Ref.~\cite{Hinchliffe} and
references therein.}\
 that has led abandoning any hope to detect it. Fortunately, the result
of this work brings some hope to remedy this issue, see the
following.

Getting back to the main stream, by using map~(\ref{RR}) into
relation~(\ref{MM}), the (total) magnetic moment in the realm of
ZBW reads
\begin{equation}\label{VV}
<\!\mbox{\boldmath$\mu$}\!>_{\rm tot}^{\rm{[NC]}_{\eta =0}}
=\frac{e}{2}<\!{\bf r}\times\mbox{\boldmath$\alpha$}\!>
+\frac{e}{4\hbar}<\!\mbox{\boldmath$\alpha$}\times({\bf
p}\times{\mbox{\boldmath$\theta$}})\!> \equiv
<\!\mbox{\boldmath$\mu$}\!>^{\rm{[C]}}+<\!\mbox{\boldmath$\mu$}\!>^{\rm{[NC]}_{\eta
=0}},
\end{equation}
where the first-part is the commutative (usual) version that has
been given in (\ref{NN}) and (\ref{OO}), and the NC part leads to
\begin{equation}\label{WW}
<\mu_3\!>^{\rm{[NC]}_{\eta =0}}
=\frac{e}{4\hbar}\left[<\!\mbox{$\alpha$}_1(\theta_1 p_3-\theta_3
p_1)+\alpha_2(\theta_2 p_3-\theta_3 p_2)\!>\right].
\end{equation}
Using the wave packet (\ref{II}), one gets
\begin{equation}\label{XX}
<\alpha_1 \theta_1 p_3\!>=2K\theta_1
\int \left [ p_1 p_3-p_1 p_3 \cos(\omega^{\rm
[zbw]}t)+p_2 p_3 \sin(\omega^{\rm
[zbw]}t) \right]f^2(p/p_{_{o}}) \,d^3 p \, ,
\end{equation}
\begin{equation}\label{XX}
<\alpha_1 \theta_3 p_1\!>=2K\theta_3 \int \left[p_1^2\left(1- \cos (\omega^{\rm
[zbw]}t)\right)+p_1 p_2 \sin(\omega^{\rm
[zbw]}t)\right]f^2(p/p_{_{o}}) \,d^3 p \, ,
\end{equation}
\begin{equation}\label{XX}
<\alpha_2\theta_2 p_3\!>=2K\theta_2\int \left[ p_2 p_3-p_2 p_3 \cos(\omega^{\rm
[zbw]}t)-p_1 p_3 \sin(\omega^{\rm
[zbw]}t) \right]f^2(p/p_{_{o}}) \,d^3 p \,
\end{equation}
and
\begin{equation}\label{XX}
<\alpha_2 \theta_3 p_2\!>=2K\theta_3 \int \left[p_2^2\left(1- \cos (\omega^{\rm
[zbw]}t)\right)-p_1 p_2 \sin(\omega^{\rm
[zbw]}t)\right]f^2(p/p_{_{o}}) \,d^3 p .
\end{equation}
Then, after taking the integrals and noticing that the terms like $\int \,d^3 p f^2(p/p_{_{o}}) p_i p_{j\neq i}$ vanish, we finally achieve
\begin{equation}\label{ZZ}
<\mu_3^{\uparrow}\!>^{\rm{[NC]}_{\eta =0}}=-\frac{e\alpha_{\rm
fsc}^2}{2\lambdabar_c}\,\theta_3[1-\cos(\omega^{\rm [zbw]} t)]\, ,
\end{equation}
 and hence
\begin{equation}
<\mu_3^{\uparrow}\!>_{\rm tot}^{\rm{[NC]}_{\eta =0}}=\frac{
e\lambdabar_c}{2 }\left\{\left[1-\left(\frac{\alpha_{\rm
fsc}}{\lambdabar_c}\right)^2\,\theta_3\right]\left[1-\cos(\omega^{\rm
[zbw]} t)\right]\right\},
\end{equation}
where, as a plausible approximation, we have considered $r_{_{o}}$
to be the Bohr radius $a_{_{\rm bohr}}=\lambdabar_c/\alpha_{\rm
fsc}$, and the definitions of $p_{_{o}}$ and $K$ have been used.
We should remind that the parameter $\mbox{\boldmath$\theta$}$
still has an insignificant value, and as it is obvious, in the
limit $\theta_3\rightarrow 0$, the NC part vanishes, as expected.

The most interesting point about (\ref{ZZ}) is that this result on
the leading-order calculations has similar structure and
numerically the same order to the result of one-loop calculations
(i.e., the second-part of relation~(\ref{UU})) in
Refs.~\cite{RiadS2000}--\cite{Eom2002}, but with an opposite sign.
Hence, in calculating the Zeeman-effect, this result suggests that
the total contributions on this part may cancel out and thus,
makes a remediation to the issue described above. However, we
should emphasis that our results are obtained by taking into
account both positive and negative energies, while the result in
the second part of relation~(\ref{UU}) has been derived just by
considering positive energy.

Incidentally, by repeating the above calculations, but with the
initial condition (\ref{bbb}) and hence its relevant wave packet,
one gets exactly the same result as relation~(\ref{ZZ}), i.e.,
$<\mu_3^{\downarrow}\!>^{\rm{[NC]}_{\eta =0}}=-e\alpha_{\rm
fsc}^2\,\theta_3[1-\cos(\omega^{\rm [zbw]} t)]/(2\lambdabar_c)$.

Up till now, we have only considered the $z$-part of the magnetic
moment, and since we have been working with the initial wave
packets with the spin in the $z$-direction, one may expect that
the other components should vanish. However, after taking the
noncommutativity effect into account and considering the relevant
wave packet, we have found that this is~not the case anymore, and
the outcomes are
\begin{equation}\label{fff}
<\mu_{1,2}^{\uparrow}\!>^{\rm{[NC]}_{\eta =0}}
=-\frac{e\alpha_{\rm fsc}^2
}{2\lambdabar_c}\left\{\theta_{1,2}[1-\cos(\omega^{\rm [zbw]}
t)]\pm {1\over 2}\,\theta_{2,1}\sin(\omega^{\rm [zbw]} t)\right\}
\end{equation}
and
\begin{equation}\label{ggg}
<\mu_{1,2}^{\downarrow}\!>^{\rm{[NC]}_{\eta =0}}
=-\frac{e\alpha_{\rm fsc}^2
}{2\lambdabar_c}\left\{\theta_{1,2}[1-\cos(\omega^{\rm [zbw]}
t)]\mp{1\over 2}\,\theta_{2,1}\sin(\omega^{\rm [zbw]} t)\right\}.
\end{equation}
Nevertheless, these results may pave the way for a new class of
experiment in detecting the NC effect on the magnetic dipole
moment. Besides, these results again are numerically in the same
order of the one-loop effect (\ref{UU}), but with an opposite
sign.

\section{Effect of Momentum Noncommutativity on ZBW}
\indent

In this section, we first consider the case of NC phase-space
background in $3+1$ dimensions with $\eta_{ij}\neq 0$ while
$\theta_{ij}=0$ for an electron in a field-free region. In this
respect, the relevant Dirac Hamiltonian has already been
derived~\cite{bertolami} to be
\begin{eqnarray}\label{hhhh}
H^{\rm{[NC]}_{\theta =0}}= c{\mbox{\boldmath$\alpha$}}\cdot{\bf
p}+ \gamma_0 m_ec^2 +{c\over
2\hbar}\left(\mbox{\boldmath$\alpha$}\times{\bf
r}\right)\cdot\mbox{\boldmath$\eta$},
\end{eqnarray}
in the SI unit, wherein it is still gauge invariant. Now,
analogous to the usual Dirac Hamiltonian for an electron in the
presence of an external magnetic field, namely
\begin{eqnarray}\label{hhh}
H_{\rm {B}}=
c{\mbox{\boldmath$\alpha$}}\cdot\mbox{\boldmath$\pi$}_{_{\rm{B}}}+
\gamma_0 m_ec^2,
\end{eqnarray}
where $\mbox{\boldmath$\pi$}_{_{\rm{B}}}={\bf p}-e{\bf A}$, one
can cast relation~(\ref{hhhh}) into the form
\begin{eqnarray}\label{iiii}
H^{\rm{[NC]}_{\theta =0}}=c\mbox{\boldmath$\alpha$}\cdot\left({\bf
p}-e{\,\mbox{\boldmath$\eta$}\times\mathbf{r}\over2
e\hbar}\right)+\gamma_0 m_ec^2 ,
\end{eqnarray}
with a corresponding generalized momentum
\begin{equation}\label{jjjj}
\mbox{\boldmath$\pi$}_{\rm{\eta}}\equiv {\bf
p}-e{\,\mbox{\boldmath$\eta$}\times\mathbf{r}\over2 e\hbar}.
\end{equation}
Also, as the vector potential due to a uniform magnetic field is
${\bf A}_{_{\rm B}}=({\bf B}\times{\bf r})/2$, one gets
\begin{equation}\label{shift}
\mathbf{A}_{\rm{\eta}}=\frac{1}{2}\frac{\,\mbox{\boldmath$\eta$}\times{\bf
r}}{e \hbar},
\end{equation}
as has been pointed out in, e.g., Ref.~\cite{NairP2001}. Such an
effect plays the role of a NC vector potential for a corresponding
similar external uniform magnetic field
\begin{equation}\label{NCmagnetic}
\mathbf{B}_{\rm{\eta}}=\frac{\mbox{\boldmath$\eta$}}{e \hbar},
\end{equation}
and in turn, if we define
$B_{\rm{\eta_{k}}}\equiv(\varepsilon_{ijk}\, B_{\rm{\eta_{ij}}})/2
$, for $B_{\rm{\eta_{k}}}=\eta_{k}/(e \hbar)$. Such an implication
resulted from $\eta_{ij}$ has been stressed in
Ref.~\cite{bertolami} and also in Ref.~\cite{farhoudi} in the
framework of gravitomagnetism. Using the bound for the NC
parameter, mentioned in Ref.~\cite{bertolami}, it yields a bound
on the magnitude of the ${\bf B}_{\eta}$-field as
$B_{\eta}\lesssim 8.6\times 10^{-14}\, {\rm T}$.

Moreover, it has been
indicated~\cite{ItzyksonZuber,holten92,Bhattacharya} that the
energy-spectrum of (\ref{hhh}), for an external magnetic field
along the $z$-direction, is
\begin{equation}\label{jjjjj}
E_{_{\rm B}}^2=m_e^2c^4+p_3^2c^2+c^2\mid\! e\!\mid  \hbar
B_3(n-l+1)-2c^2\, e B_3 s_{3},
\end{equation}
where $s_{3}=\pm\hbar/2$, $n$ is a non-negative integer and
$L_3=\hbar l$, while $l$ is restricted to values $l=-n, -n+2,
\cdots, n-2, n$. Also, for electron, the relativistic form
(\ref{jjjjj}) can be rewritten as
\begin{equation}\label{jjj}
E_{_{\rm B}}^2=m_e^2c^4+p_3^2c^2+2 k c^2\mid\! e\!\mid  \hbar B_3,
\end{equation}
which is called the Landau energy-levels~\cite{Bhattacharya}.
Thus, there exists a two-fold degeneracy in the solutions, i.e.,
spin-up with $n-l=2( k-1)$ and spin-down with $n-l=2 k$. However,
the case $k=0$ (i.e., the zero Landau-level) is a specific case,
for $(n-l)$ cannot be negative. That is, in this specific case,
for electron, only the spin-down solution exists, and in the same
vein, for positron, only the spin-up solution is allowed. Hence,
for the zero Landau-level, the degeneracy of the spectrum is
lifted.

Analogously, in the case of NC geometry, without loss of
generality, when only $B_{\rm{\eta_{3}}}\neq 0$, its corresponding
energy-spectrum (like relation (\ref{jjjjj})) becomes
\begin{equation}\label{jjjE}
E_{\rm{ \eta}}^2=m_e^2c^4+p_3^2c^2+c^2\mid\! e\!\mid \hbar
B_{\rm{\eta_{3}}}(n-l+1)-2c^2\, e B_{\rm{\eta_{3}}} s_{3},
\end{equation}
or also
\begin{equation}\label{jjjEA}
E_{\rm {\eta}}^2=m_e^2c^4+p_3^2c^2+2 k c^2\mid\! e\!\mid \hbar
B_{\rm{\eta_{3}}}.
\end{equation}
However, as the ${\bf B}_{\eta}$-field, in relation
(\ref{NCmagnetic}), depends on the electric charge, the
energy-levels are independent of the sign of electric charge. More
interestingly, in the NC region where only $\eta_{ij} \neq 0$ with
no external field, solutions of the Dirac equation for an electron
in the presence of an external magnetic field hold while the ${\bf
B}_{\eta}$-field has been replaced instead of external magnetic
field.

Now, we propose to apply the above result for the issue of an
electron in external magnetic field in $2+1$ dimensions, the
domain that more attentions have been paid to in recent decades.
This issue has already been considered both in commutative, e.g.,
Refs.~\cite{Zawadzki2011,Zawadzki}, and NC, e.g.,
Refs.~\cite{Hassanabadi,bertolami2+1}, contexts, where the ZBW
effect in graphene has been studied in the former references.

To proceed and specify the characteristic feature of the ZBW in
the momentum noncommutativity, we preferably employ the results of
Refs.~\cite{Zawadzki2011,Zawadzki}, wherein the Hamiltonian for
electrons and holes, at the $K_1$ point for a monolayer graphene
in the presence of an external magnetic field along the z-axis,
has been given to be
\begin{equation}\label{2+1Hamilton}
H = v_{_F}\left(\sigma_1 \pi_1+\sigma_2\pi_2\right),
\end{equation}
where $v_{_{F}}\approx 1\times {10}^8\, {\rm cm/s}$ is the Fermi
velocity, and hereinafter, we will replace
$\mbox{\boldmath$\pi$}_{\eta}$ instead of $\mbox{\boldmath$\pi$}$.
Then, without loss of generality, by choosing the NC vector
potential as the Landau gauge
$A_{\rm{\eta_{2}}}=0=A_{\rm{\eta_{3}}}$ and $A_{\rm{\eta_{1}}}=-y
B_{\eta} $, the energy eigenvalues are $E_{sm
}=s\hbar\Omega\sqrt{m}$, where angular frequency
$\Omega\equiv\sqrt{2\mid\! e B_{\eta}\!\mid\!/\hbar}\,
v_{_{F}}=\sqrt{2 \eta}\, v_{_{F}}/\hbar$, oscillator numbers
$m=0,1,\ldots$, and energy branch $s=\pm 1$ stands for the
conduction and valence bands, respectively.

The average of velocity operator, $ \dot{r}_i(t)$ for $i=\{1,2\}$,
through an arbitrary two component function, say $f$, in the
Heisenberg picture, is
\begin{equation} \label{V_vi(t)}
\langle \dot{r}_i(t) \rangle = \sum_{\rm m',m} e^{iE_{\rm
m'}t/\hbar}\langle f|{\rm m'}\rangle \langle {\rm
m'}|\dot{r}_i(0)|\rm m \rangle \langle {\rm m}|f\rangle
e^{-iE_{\rm m}t/\hbar},
\end{equation}
where from the Hamiltonian equation $\dot{r}_i(0)=\partial
H/\partial\pi_i$. Also, $|{\rm m}\rangle$ is eigenstate solution
of Hamiltonian~(\ref{2+1Hamilton}) that, in the form of harmonic
oscillator function labeled by three quantum numbers, has been
derived to be~\cite{Zawadzki2011,Zawadzki}
\begin{equation} \label{H_nskx}
|{\rm m}\rangle \equiv |m k_xs\rangle = \frac{e^{ik_x
x}}{\sqrt{4\pi}} \left(\begin{array}{c} -s|m-1\rangle
\\|m\rangle
\end{array}\right),
\end{equation}
where $|m\rangle$ is $m$-th state of the harmonic oscillator,
hence the summation in relation (\ref{V_vi(t)}) should go over all
the quantum numbers, i.e. $\sum_{\rm m',m}\rightarrow \int\int
dk'_x dk_x\sum_{m',m}\sum_{s',s}$.

To get a better and prompt insight on the nature of the ZBW in
graphene, we choose the arbitrary function $f(x,y)$ in the form of
a circular Gaussian wave packet
\begin{equation} \label{Gauss_f}
f(x,y) = \frac{1}{\sqrt{\pi }\,\ell}
e^{-\frac{x^2+y^2}{2\ell^2}-ik_{0
x}x} \left(\begin{array}{c} u \\
d\end{array}\right),
\end{equation}
where $p_{0 x}=\hbar k_{0 x}$ is an initial nonvanishing momentum,
$\ell $ is width of the wave packet, and $u^2+d^2=1$, however we
assign $u=d=1/\sqrt{2}$. Hence, the average of velocity components
become
\begin{eqnarray}\label{V_vx(t)}
\langle  \dot{r}_1(t) \rangle\!\!\!\!\! &=&\!\!\!\!\!  v_{_{F}}
\sum_{m=0}^{\infty} \left[\alpha_m^+
\cos(\omega_m^{\rm [cyc]}t) + \alpha_m^-\cos(\omega_m^{\rm [zbw]} t)\right], \\
\langle \dot{r}_2(t) \rangle\!\!\!\!\! &=&\!\!\!\!\!
v_{_{F}}\sum_{m=0}^{\infty}\left[ \beta_m^+\sin(\omega_m^{\rm
[cyc]}t) + \beta_m^-\sin(\omega_m^{\rm [zbw]} t)\right],
\label{V_vy(t)}
\end{eqnarray}
where cyclotron frequency $\omega_m^{\rm
[cyc]}\!\equiv\Omega\left(\sqrt{m+1}-\sqrt{m}\right)$,
$\omega_m^{\rm
[zbw]}\!\equiv\Omega\left(\sqrt{m+1}+\sqrt{m}\right)$,
$\alpha_m^{\pm}\!\equiv 2\left(V_{m,m} \pm V_{m-1,m+1}\right)$,
$\beta_m^{\pm}\equiv -2(V_{m,m+1}\pm V_{m,m-1})$ with $
V_{m,m'}\equiv \int F_{m}(k_x) F_{m'}(k_x) dk_x $
while~\cite{Zawadzki}
\begin{equation} \label{Gauss_Fn_wyn}
F_m(k_x)= \frac{\ell\
\sqrt{L(L^2-\ell^2)^m}}{\sqrt{2^{m+1}(L^2+\ell^2)^{m+1}
m!\sqrt{\pi} }} e^{-\frac{1}{2}\ell^2(k_x-k_{0x})^2}
e^{-\frac{k_x^2L^4}{2(L^2+\ell^2)}}\ {\rm H}_m(k_x g).
\end{equation}
Here, $ {\rm H}_m(k_x g)$ is the Hermit polynomials, the magnetic
radius $L=\sqrt{\hbar/(\mid\! e B_{\eta}\!\mid
)}=\hbar/\sqrt{\eta}$ and $g\equiv L^3/\sqrt{L^4-\ell^4}$. In
addition, from relations (\ref{V_vx(t)}) and (\ref{V_vy(t)}), we
easily get the expectation value of position as
\begin{eqnarray} \label{V_vx(t)2}
\langle  r_1(t) \rangle\!\!\!\!\! &=&\!\!\!\!\!
v_{_{F}}\sum_{m=0}^{\infty} \left[\frac{\alpha_m^+}{\omega_m^{\rm
[cyc]}}
\sin(\omega_m^{\rm [cyc]}t) + \frac{\alpha_m^-}{\omega_m^{\rm [zbw]} }\sin(\omega_m^{\rm [zbw]} t)\right],\\
\langle r_2(t) \rangle\!\!\!\!\! &=&\!\!\!\!\! -
v_{_{F}}\sum_{m=0}^{\infty} \left[\frac{\beta_m^+}{\omega_m^{\rm
[cyc]}}\cos(\omega_m^{\rm [cyc]}t) +
\frac{\beta_m^-}{\omega_m^{\rm [zbw]} }\cos(\omega_m^{\rm [zbw]}
t)\right]. \label{V_vy(t)2}
\end{eqnarray}

Therefore, in comparison with the (usual) commutative scenario,
relations (\ref{V_vx(t)}), (\ref{V_vy(t)}), (\ref{V_vx(t)2}) and
(\ref{V_vy(t)2}) indicate that, due to momentum noncommutativity,
ZBW oscillations are permanent with many frequencies while
accompanied with cyclotron frequencies as well. In the case of
usual external magnetic field, to get a better realization in
application, one can assume the magnetic radius $L$ being equal to
the width of wave pack $\ell$, wherein the only surviving term in
those relations will be the zero Landau-level. However, in the
case of ${\bf B}_{\eta}$-field, as $L\gtrsim 8.7\, {\rm cm}$, this
presuppose diverts the issue away from the quantum realm.

Nevertheless, to realize the role of a specific Landau-level in
the ZBW motion, we should confine ourselves to a specific value of
$m $, say $m_{_{0}}$, thus in this case, the motion in the
$xy$-plane is
\begin{eqnarray}\label{V_vx(t)3}
\langle  r_1(t) \rangle_{m_{0}}\!\!\!\!\! &=&\!\!\!\!\!
v_{_{F}}\left[\frac{\alpha_{m_{0}} ^+}{\omega_{m_{0}}^{\rm [cyc]}}
\sin(\omega_{m_{0}}^{\rm [cyc]}t) + \frac{\alpha_{m_{0}} ^-}
{\omega_{m_{0}}^{\rm [zbw]} }\sin(\omega_{m_{0}} ^{\rm [zbw]} t)\right], \\
\langle r_2(t) \rangle_{m_{0}}\!\!\!\!\! &=&\!\!\!\!\! - v_{_{F}}
\left[\frac{\beta_{m_{0}}^+}{\omega_{m_{0}}^{\rm
[cyc]}}\cos(\omega_{m_{0}}^{\rm [cyc]}t) + \frac{\beta_{m_{0}}
^-}{\omega_{m_{0}}^{\rm [zbw]} }\cos(\omega_{m_{0}}^{\rm [zbw]}
t)\right]. \label{V_vy(t)3}
\end{eqnarray}
Still to extract a physical picture, let us concentrate on the
zero Landau-level. In this particular case, $\omega_0^{\rm
[zbw]}=\Omega =\omega_0^{\rm [cyc]}$ wherein $\Omega\lesssim
1.6\times 10^7\, {\rm s^{-1}}$, and
\begin{eqnarray}\label{r1zero}
\langle  r_1(t) \rangle_{0}\!\!\!\!\! &=&\!\!\!\!\!
\frac{4 v_{_{F}}V_{0,0}} {\omega_0^{\rm [zbw]}  }\sin(\omega_0^{\rm [zbw]}  t),\\
\langle r_2(t) \rangle_{0}\!\!\!\!\! &=&\!\!\!\!\! \frac{ 4
v_{_{F}}V_{0,1}} {\omega_0^{\rm [zbw]}  }\cos(\omega_0^{\rm [zbw]}
t),\label{r2zero}
\end{eqnarray}
where
$$
V_{0,0}=\frac{L\,\ell^2}{2\sqrt{(L^2+\ell^2)(L^4+L^2\ell^2+\ell^4)}}
\exp\left[{\frac{\ell^2k^2_{0x}(\ell^2-1)(L^2+\ell^2)}{L^4+L^2\ell^2+\ell^4}}\right]
$$
and
$V_{0,1}=\left[L^3\ell^2k_{0x}\sqrt{2}/(L^4+L^2\ell^2+\ell^4)\right]V_{0,0}$.
These results clearly indicate the rotational nature of the ZBW
motion for the zero Landau-level in the presence of the $\bf
B_{\eta}$-field. However, considering the approximate values of
$L$ and $\ell$, the amplitudes of (\ref{r1zero}) and
(\ref{r2zero}) are roughly
$\left(\ell^2\sqrt{\eta}/\hbar\right)\exp{\left[-(\ell\,
k_{0x}/\hbar)^2\eta\right]}$ and $\left(\ell^4
k_{0x}\eta/\hbar^2\right)\times\exp{\left[-(\ell\,
k_{0x}/\hbar)^2\eta\right]}$, respectively. Nevertheless, it is
interesting to note that, if the ZBW part is absent, the amplitude
of the motion will be halved, which gives an idea for a probable
experimental check.

\section{Conclusions}
\indent

First, we reviewed the ZBW in the commutative (usual) spacetime.
Then, we have considered the effects of the space and momentum
noncommutativity separately on the ZBW phenomenon. In both cases,
the states of positive and negative energy Dirac electrons have
been used to derive the relevant wave packets, by which we have
calculated the effects of the NC parameters on the ZBW. In the
space noncommutativity scenario in $3+1$ dimensions, it has been
expressed that, due to the conservation of momentum, the Fourier
decomposition of the expectation value of position does~not
change. However, we have represented that the electron magnetic
moment receives correction in the same manner as in the
commutative version and is proportional to the NC parameter, but
spin-independent and with an insignificant value. Besides, this
correction is due to the leading-order calculations, whereas, in
the literature, it has been indicated that such a correction would
appear only in the one-loop level. And more interestingly, we have
found that this correction has similar structure and numerically
the same order to the result of one-loop calculations, but with an
opposite sign. Hence, when calculating the Zeeman-effect of an
atom, the contributions of these two parts may cancel out each
other and therefore, for large lumps of masses, the net result due
to the interaction of the NC part of the magnetic moment with an
external magnetic field in the Zeeman-effect remains undetectable.
Moreover, we have shown that the $x$- and $y$-components of the
electron magnetic dipole moment, contrary to the commutative case,
are non-zero and proportional to the NC parameter. This result may
pave the way for a new class of experiment in detecting the NC
effect on the magnetic dipole moment. Besides, their corrections
again are numerically in the same order of the one-loop effect,
but with an opposite sign.

In the momentum noncommutativity scenario, as it has been
expressed in the literature, this effect is equivalent to
introducing an external uniform magnetic field into the Dirac
equation. Hence, one expects that the solutions of the Dirac
equation should change accordingly. Indeed, in this case, we have
shown that the energy-spectrum and also the solutions of the Dirac
equation are affected in both $3+1$ and $2+1$ dimensions. In this
respect, we mostly focused on $2+1$ dimensions for the specific
case of graphene, where we have taken the solutions to figure out
the effect of the NC parameter and its relevant field on the ZBW.
We have indicated that the resulted NC field affects the motion of
electrons in the zero Landau-level not only via a cyclotron
motion, that is usually expected from any external magnetic field,
but also through the ZBW motion which doubles the amplitude of the
rotation in the plane perpendicular to the direction of the field.
In fact, this is a hallmark of the ZBW in graphene that provides a
promising way to be tested experimentally, however the correction
is proportional to the NC parameter with an insignificant value.
Generally, in comparing with the (usual) commutative scenario,
even though the structure of the corresponding NC field is the
same as an external uniform magnetic field and the results show
the rotational nature of the ZBW motion, but it treats with many
frequencies of the ZBW while accompanied with many cyclotron
frequencies as well.

\setcounter{equation}{0}
\renewcommand{\theequation}{A.\arabic{equation}}

\section*{Acknowledgements}
\indent

This work has been supported by National Elites Foundation of Iran
via M.Z.A., and we also thank the Research Council of Shahid
Beheshti University. Besides, M.Z.A. appreciates prof. H. Arfaei
for helpful discussions.

%
\end{document}